\documentclass[twocolumn,showpacs,preprintnumbers,amsmath,
amssymb]{revtex4}
\usepackage{graphicx}
\usepackage{dcolumn}
\usepackage{bm}
\usepackage{tensor} 
\begin{document}
 
\title{Optical counterpart of Foucault pendulum.}
 
\author{A.Yu.Okulov}
\email{alexey.okulov@gmail.com}
\homepage{https://sites.google.com/site/okulovalexey }
\affiliation{Russian Academy of Sciences, 119991,  Moscow, 
Russian Federation.}
 
\date{\today}
 
\begin{abstract}
{The twin beam vortex interferometer with phase-conjugating 
mirror is analyzed in rotating reference 
frame.Circular motion of the interference pattern occurs 
due to exchange of the angular momenta between photons and interferometer.  
Using the concept of the $ideal$ 
phase-conjugating mirror it is shown that motion of the 
helical interference pattern of the interacting vortex photons 
with topological charge $\ell$ may be used for the detection 
of the slow rotations. 
The higher density of interference fringes may improve the sensitivity 
by factor containing $2 \ell$ compared to the conventional 
Michelson interferometry.}
\end{abstract}

\pacs{42.50.Tx 42.65.Hw 42.50.Lc 04.80.Nn }
\maketitle

\vspace{1cm}

\section {Introduction}

The rotation of the Earth was a one of the 
most controversial issues of natural philosophy during 
centuries in transition from Medieval period to Renaissances and 
afterwards. 
The invention of Foucault pendulum \cite {Foucault:1852} 
and  mechanical gyro  realized 
by Johann Bohnenberger in 1817 
did not stopped these  
controversies but ignited further studies of 
the Earth motion stimulated by navigation needs. 
The necessity to check the hypothesis of the "ether wind" have 
led to construction of the highly effective 
optical instruments \cite {Born_Wolf:1972}. 
Michelson interferometer gave possibility 
to detect the small displacements 
and star's dimensions \cite {Michelson:1904}. 
Saqnac gyro proved to be highly sensitive to phase lag 
between counterpropagating 
waves caused by rotation of the reference frame
\cite {Sagnac:1913}. Nowadays Maxwell electrodynamics and 
Einstein relativity explain well the Sagnac effect 
which is in the heart 
of the widespread rotation sensors, technically 
implemented as a passive fiber gyroscopes and the active 
laser gyros \cite {Scully:1997}. 
In the LIGO project \cite {LIGO:2016,Pustovoit:1962} the 
Michelson interferometry with ultra long arms 
and ultra narrow linewidth laser sources have become conventional 
instrument in the gravitational waves search \cite {Braginsky:2009}. 
\begin{figure}
\center{ \includegraphics[width=4.0 cm] {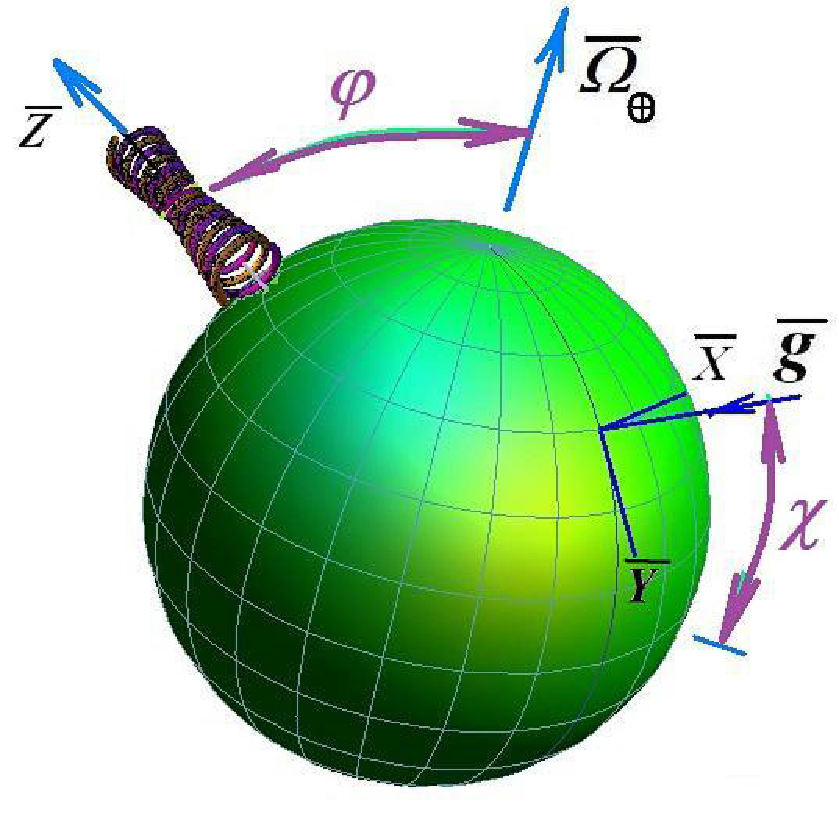}}
\caption { (Color online) Geometric phase $\alpha(t)$ 
acquired by Foucault pendulum and vortex interferometer with 
PCM. For the angle $\phi$ 
between $\ell$-charged 
vortex rotation $\vec Z$ axis 
and frame rotation axis $\vec \Omega_{\oplus}$ the 
geometric phase is 
$\alpha_{pc}(t)= - 2 \ell |\Omega_{\oplus}| t  \cos \phi$ 
(for $\xi = 0$). 
Local 
gravity acceleration $\vec {\bf g}$ indicates 
location of Foucault pendulum whose
swing plane is rotated by 
Coriolis force to angle 
$\alpha_F(t)= - |\Omega_{\oplus}| t \sin \chi$. $X$,$Y$,$-\vec g$ 
are local 
coordinates for pendulum, $\chi$ is latitude. The angle 
$ \phi$ is in meridional plane and $\xi$ 
is out of meridional plane angle  
($\xi $ is set to zero hereafter for brevity without 
loss of rigor).}
\label{fig.1}
\end{figure}

In current work we analyze the new principle of 
the reference frame rotation detection 
based upon angular Doppler effect for photons 
\cite {Okulov:2012josa} and wavetrain dislocations 
\cite {Berry:1974} (fig.1). 
The rotations of optical quanta are 
different from a classical mechanical top. 
Compared to classical mechanics 
the angular momentum projection on given axis $\vec Z$ 
may have only discrete values proportional to the 
Plank's constant $\hbar$ \cite {Allen:1992,Barnett:2002}. 
The other feature of the optical vortex interferometry 
is robustness of the vortex beams \cite {Boyd:2014, Soskin:2000} 
with respect to the irregularities in optical path \cite {Gbur:2008}.
This is due to conservation of the orbital 
angular momentum (OAM) projection 
${\mathcal L}_z=\pm \ell \hbar$ in a free space
\cite {Gbur:2016} for the optical vortices 
\cite {Padgett:2009,Soskin:2008} with winding 
number $\ell$ . On the other hand 
rotations of the optical elements in interferometer 
affect propagating photons \cite {Dholakia:2002D}. 
This gives experimental 
possibility to detect the angular velocity $\vec \Omega_{\oplus} $ 
of the reference frame 
with optical vortex interferometer and angular Doppler effect. 
The key feature of the proposed phase-conjugating vortex 
interferometer (PCVI) is the 
usage of wavefront reversing mirror (PCM) 
\cite {Basov:1980} which alters direction of the photon's 
angular momentum \cite {Okulov:2008}. 

In comparison with previous considerations that use the concept  
of geometric phases and 
analogies with Foucault pendulum and Coriolis force 
experienced by photons 
\cite {Pancharatnam:1956,Berry:1984,Berry:1987,Tomita_Chiao:1986,
Senthilkumaran:2000,Culshaw:2000,Ruchi:2017} current approach demonstrates 
visual similarity of the circular interference pattern rotating due to angular  
Doppler shift and $seemingly$ circular motion of stop points of Foucault 
pendulum bob (fig.2). In both cases the observable motion is induced via nontrivial 
angular momentum changes in noninertial reference frame. Noteworthy the  
exact solutions for angular velocity of rotation exist in 
explicit form in both cases.

The  angular momentum of photons is 
not changed when observed  
from reference frames rotating with different angular 
velocities \cite {Landau:1982}. But when rotating photons 
pass the rotating medium the 
angular momentum does change and the medium experiences 
recoil \cite {Okulov:2012josa,Fiona:2014}. 
This happens when vortex photons propagate in the 
interferometer placed upon a surface of the rotating rigid 
sphere \cite {Bliokh:2008} (fig.1). 
As a result of the such spin-orbital interaction the 
carrier frequency $\omega$ of the vortex photon is 
shifted as follows:
\begin{equation}
\label {Coriolis shift} 
\delta \omega (t) = -  \vec {\mathcal L}(t) \cdot \vec \Omega_{\oplus}/ \hbar = 
- \ell \vec {Z}(t) \cdot \vec \Omega_{\oplus}.
\end{equation}  
When interferometer is placed 
somewhere in equatorial plane  with $\chi=0$ or at the 
Pole with $\pm \chi=90^o$  (where $\chi$ is 
geographical latitude) $\delta \omega$ 
is just rotational Doppler shift which has maximal value 
$\delta \omega =  \mp \Omega_{\oplus} {\mathcal L}{_z} /\hbar$ at $\phi= 0$ 
and $\delta \omega = 0$ for $\phi= \pi/2$ .
For arbitrary location of PCVI spin-orbit interaction  
leads to Coriolis frequency shift of photon 
$\delta \omega (t) = 
- \vec {\mathcal L}(t) \cdot \vec \Omega_{\oplus} /{\hbar} $. 
In the same way the conventional
Foucault pendulum operates due to the geometric phase 
$\alpha = -2 \pi \sin \chi$ \cite {Landau:1976} acquired via 
transport of rotating top along a closed trajectory, 
where $\phi=\pi/2 - \chi$ is the angle between rotation axis $\vec \Omega_{\oplus}$ 
and angular momentum $\vec {\mathcal L_F}(t)$, $\chi$ is the 
geographical latitude (fig.1).

The paper is organized as follows. In section II the geometry 
of PCVI is described as an extension of the Beth spin 
angular momentum detection experiment \cite {Beth:1936}, 
in the section III the interference patterns 
in both arms of the vortex Michelson interferometer 
are analyzed taking into account the imperfect 
coherence $\gamma$ of the laser source, 
in the section IV the geometric phase 
shift $\alpha = \int \delta \omega d \tau$ 
is obtained from the space-time symmetries, in the  
section V the effects of the phase noise of laser source 
\cite{Townes:1958} and PCM noise \cite{Gaeta:1988} 
are evaluated taking into account the possibility of 
quantum noise reduction with topologically charged LG-beams
\cite{Lugiato:1997}, in section VI the results 
are summarized.  
\section {Configuration of the vortex Michelson interferometer}
Let us begin with the classical experiment on the 
optical angular momentum measurement performed by 
Beth in 1936 \cite {Beth:1936}. 
The circularly polarized light with 
angular momentum $\pm \hbar$ per photon had been transmitted 
through the $\lambda /2$ plate  
suspended on the quartz wire.  
Such transparent plates are made 
from $anisotropic$ material 
(quartz) which changes the 
angular momentum of the each photon to the opposite 
one $\mp \hbar$ during passage through the plate. 
In accordance with the second Newton's 
law and the angular momentum $\vec {\mathcal L}$ 
conservation the plate experienced the 
torque 
$\vec T=\Delta {\vec {\mathcal L}}/ \Delta t$, where 
$\Delta {\vec {\mathcal L}}$ is the angular momentum change during 
time interval $\Delta t$. The torque occurs because of 
the $noncollinearity$ of the 
electric field vector of light $\vec E$ and 
macroscopic polarization $\vec P dV$ 
(dipole moment of the volume $dV$) inside the birefringent plate. This 
noncollinearity is manifestation of an anisotropy 
of the $\lambda/2$ plate which makes the vector product 
$\vec E \times \vec P$ nonzero. The arising torque is   
$\vec T=\epsilon_0 {\int}(\vec P \times \vec E)dV
\cong 2 \cdot I \cdot \pi D_p^2 /{\omega_{f,b}}$, 
where 3D integral is calculated over 
the plate volume, $I$ is the light intensity, $D_p$ is 
diameter of plate and $\omega_{f,b}$ are the  
carrier frequencies of the light waves which travel in the forward (f) or 
backward(b) direction of $Z-axis$. 
Thereby the suspending 
wire had been twisted and a certain deflection of the plate 
from the equilibrium 
position had been detected. In order to enhance the torque 
Beth reflected the 
light backwardly by a traditional 
metallic mirror. The important feature of his 
setup is an additional $\lambda /4$ plate placed near 
mirror $to$ $alter$ $spin$ $angular$ $momentum$ 
to double the optical torque via return passage through the 
suspended $\lambda /2$ plant. Without $\lambda /4$ plate the 
algebraic sum of torques on the suspended plate would be zero. 
In fact $\lambda /4$ plate performed the phase-conjugation 
of reflected wave and the $spin {\:}
angular {\:}momentum$ of photons had been reversed. 

Our proposal is to 
replace the traditional mirror by wavefront 
reversing mirror \cite {Zeldovich:1985}  
which alters $orbital {\:} angular {\:} momentum $ (OAM)
of photons \cite {Okulov:2008}, replace $\lambda /2$ plate 
by a sequence of the $N$ image altering elements 
(alike Dove prism) \cite {Soskin:2008} and to 
use a higher-order optical vortices with angular 
momentum $\pm \ell \hbar$ per photon \cite {Barnett:2002} 
instead of circularily polarized light whose AM is just $\pm \hbar$. 
The possible phase-conjugation mirrors for this purpose might be  
the photorefractive crystals \cite {Saffman:1996} or the 
static 3D holograms \cite {Polyansky:2005}. 
The else opportunity is nondegenerate four-wave mixing 
in an alkali atomic vapors where efficient phase conjugated 
reflection from $10^5$ atoms in thermal cloud had been 
reported \cite {Tabosa:1999}. 
The other tool 
for the phase conjugation of the $\pm \ell \hbar$ optical vortices 
is in multiple reflections from conventional mirrors 
in the Sagnac loop interferometer \cite {Okulov:2010josa}. 
Thus by virtue of the phase conjugating mirror 
Beth's torsion pendulum setup is transformed 
into $vortex{\:} {\:}interferometric$ setup realized in 
Denz group \cite {Woerdemann:2009} (fig.2). Instead of the altering 
the $spin$ component of photon's angular momentum, 
the alternation of the $orbital {\:} angular {\:}momentum$ 
had been realized in this setup with commonly available 
optical components \cite {Padgett:1998}. 

The interference pattern between the beamsplitter BS and PCM 
which arise due 
to the reversed orbital angular momentum of the backwardly  
reflected phase conjugated 
replica $E_b(t,z,r,\theta)=E_f^{*}(t,z,r,\theta)$ has 
a nontrivial geometry. In contrast to the 
speckle patterns composed of vortex-antivortex pairs 
\cite {Okulov:2009} and the complex regular images generated 
by vortex arrays\cite {Volostnikov:1989} this 
isolated vortex pattern is composed of the $2\ell$ 
mutually embedded helices ({fig.2}) 
\cite {Okulov:2008J,Woerdemann:2012}: 
\begin{eqnarray}
\label{Archimedean_inter_patt}
{|\vec E|^2}={|E_f + E_b|^2}\cong
{I}{(z,r,\theta,t )} \sim [ 1 + \gamma [2(L_{_{PCM}}-z)] \cdot  {\:}
&& \nonumber \\
cos[ {\:} (\omega_f-\omega_b ) t - (k_f+k_b) z + 
{2\ell}{\:}\theta ]] \cdot {\:}
&& \nonumber \\
{({r}/{D_0})^{2|\ell|}}{\:}
exp{\:}[{\:}-
{\frac {2 r^2}{{D_0}^2(1+z^2/({k_{(f,b)}}^2 {D_0}^4))}}]
{\:}{\:},{\:}{\:}
\end{eqnarray} 
where the cylindrical coordinates $(z,r,\theta,t)$ 
are used, $k_{f,b}$ are the wavenumbers of the 
$E_f$ and $E_b$ respectively, 
${I}{(z,r,\theta,t )}$ is the light intensity distribution of 
the $2 \ell$ intertwined helices, $\gamma [2(L_{_{PCM}}-z)]$ is 
the temporal correlation function of the laser beam which vanishes 
when $2(L_{_{PCM}}-z) > L_{coh}$, $ L_{coh}$ is coherence 
length of the laser source, $L_{_{PCM}}$ is length of the PCM arm, 
$D_0$ is the radius of 
$LG_{0\ell}$ vortex. Apart from spirality the formula 
(\ref {Archimedean_inter_patt}) describes 
$synchronous{\:}rotation{\:}$ of the all $2 \ell$ helices 
around the propagation axis $z$ with angular velocity 
$(\omega_f-\omega_b)/2\ell  $ \cite {Okulov:2008}. The rotation 
appears   
when the carrier frequencies of the forward $E_f$ and backward waves 
$E_b$ are different. 
The electric field envelopes were taken above in the form 
of Laguerre-Gaussian beams (LG) \cite {Okulov:2008}: 

\begin{eqnarray}
\label{pump_LG}
{{ E}_{(f,b)}(\vec r,t)}
\sim {\frac {\exp [ -i \omega_{(f,b)}t 
\pm i k_{(f,b)} z \pm i{\ell}\theta + i \Theta_{(f,b)}(t)]} 
{ {(1{+}iz/z_R)}} } 
&& \nonumber \\
\times {{\ E}_{(f,b)}^0}{({r}/{D_0})^{|\ell|}}
\exp  [  - 
{\frac {r^2}{2{D_0}^2(1{+}iz/z_R)}} ], 
z_R=k_{(f,b)} {D_0}^2{\:} {\:}.
\end{eqnarray}
Alternatively the Bessel beam optical vortices
may be considered  \cite {Sepulveda:2009}:
\begin{eqnarray}
\label{pump_Bessel}
{{ E}_{(f,b)}(\vec r,t)}
\sim{{\ E}_{(f,b)}^0} \cdot J_m(\kappa r)
&& \nonumber \\
{  \exp [ -i \omega_{(f,b)}t 
\pm i k_{(f,b)}z \pm i{\ell}\theta + i \Theta_{(f,b)}(t)]}. 
\end{eqnarray}
In both cases the random variations of the phases of 
vortex waves (phase diffusion) $\Theta_{(f,b)}(t)$ are  
induced by the finite laser 
linewidth with coherence time $\tau_{coh}=L_{coh}/c$ \cite {Siegman:1986}. 
Phase diffusion $\Theta_{(f,b)}(t)$ leads to the diminished 
visibility $\gamma [2(L_{_{PCM}}-z)]$ for nonzero path difference. 

We will consider the frequency splitting 
induced by the angular momentum transfer from photons  
to $rotating$ interferometer or vice versa. 
Due to OAM exchange 
photons may acquire the energy from rotating interferometer 
components or deliver it to interferometer.

\begin{figure} 
\center{ \includegraphics[width=7.0 cm]
{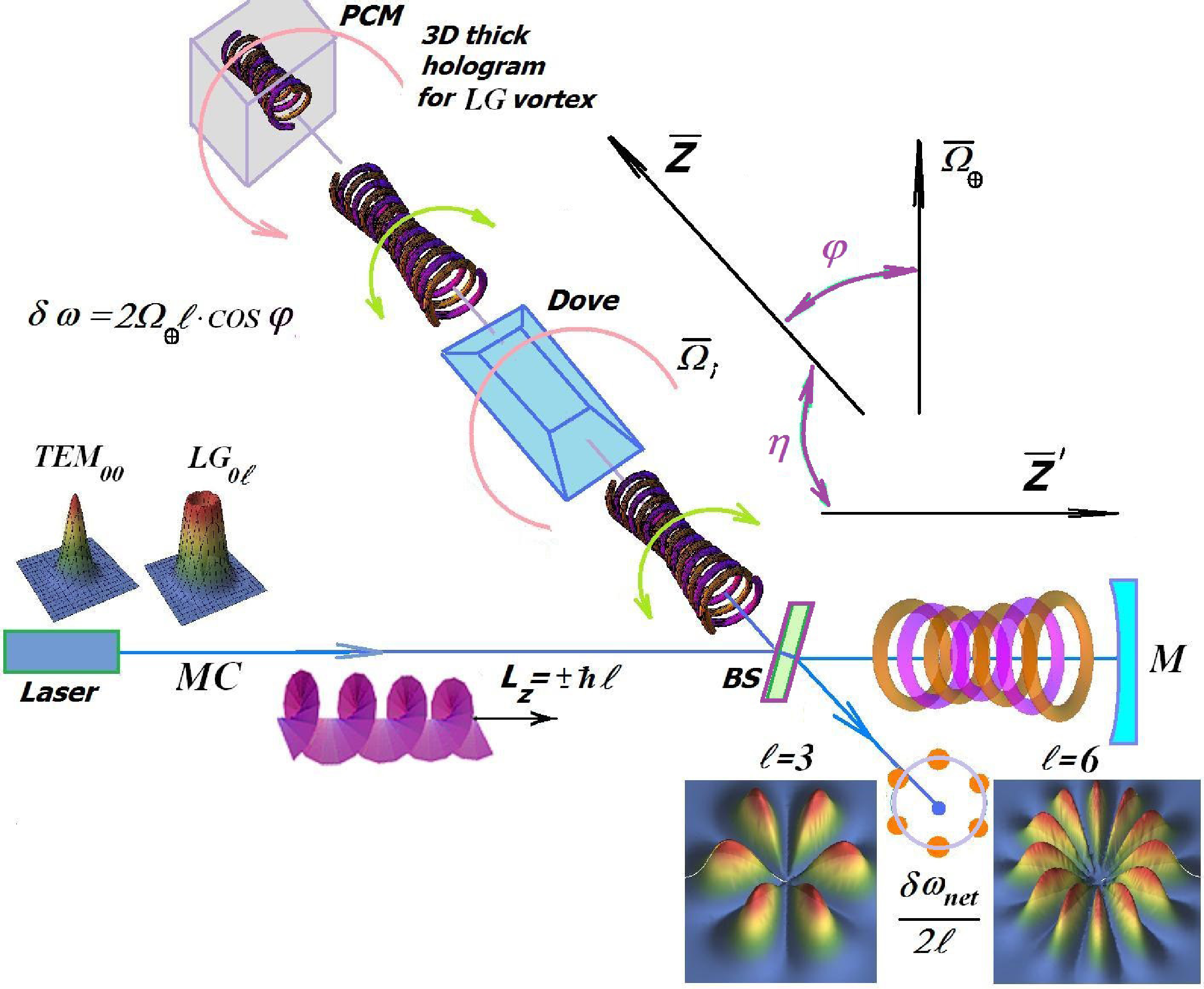}}
\caption{ (Color online) Phase-conjugating vortex interferometer 
PCVI with topological charge $\ell$ \cite{Woerdemann:2009}
aligned along the axis $\vec Z{\:}$. 
The azimuthal interference fringes for $\ell=3$ 
and for $\ell=6$ are shown. 
Single spatial mode $TEM_{00}$ laser 
output is transformed by mode converter $MC$ 
in optical vortex with topological charge $\ell=1,3,6,etc$. 
The $N$ counter-rotating with respect to each other Dove prisms 
with angular velocities $\vec \Omega_i$ (only one shown) 
and PC mirror rotating with angular velocity $\vec \Omega_{\oplus}$
alter the photon's angular 
momentum thereby the angular frequency shift 
$\delta \omega_{net}$ appears. 
Angle $\phi$ 
is a tilt of $\vec Z$ axis  
to the frame rotation axis $\vec \Omega_{\oplus}$.
The angle $\eta$ 
between $\vec Z$ and $\vec Z^{'}$ axes affects interference pattern 
between "helical" and "toroidal" 
arms of Michelson PCVI. Response of interferometer 
reaches maximum when $\eta$ tends to zero.} 
\label{fig.2}
\end{figure}

\section{Spatial patterns due to exchange of rotations
between photons and interferometer}

The mutual exchange of the energy and angular momentum between 
photons and Mach-Zehnder interferometer had been reported 
by Dholakia group in 2002 yet \cite {Dholakia:2002} where 
the the interference patterns revolving with 
Hz-order frequencies were recorded. In essence there is 
no difference whether a single 
element rotates say Dove-prism \cite {Padgett:1998} 
or $\lambda/2 $ plate  
\cite {Dholakia:2002} or entire 
interferometric setup is rotated as a whole. 
In all these cases 
the rotational Doppler shift $\delta \omega$ 
occurs due to 
the angular momentum exchange between photons and setup. 
The phase-conjugating mirror will substantially 
simplify the implementation of such $sub-Hz$ 
rotation sensor because of the self-adjustment property of 
the PCM \cite {Basov:1980}. The perfect match of 
amplitudes and phases of forward and backward waves achieved 
in the Woerdemann-Alpmann-Denz photorefractive interferometer 
setup \cite {Woerdemann:2009} have resulted in a 
remarkable $two-spot$ output pattern obtained by virtue 
of the beamsplitter BS placed at the entrance of interferometer
(fig.2). Two-spot output of this vortex phase-conjugating 
interferometer \cite {Woerdemann:2009} 
is result of the usage of the $\ell=1$ optical vortex 
($LG_{01}$) laser beams. For the higher angular momenta of photons 
$\ell \hbar$ the output interference pattern have $2\ell$ spots:
\begin{eqnarray}
\label{pines_inter_patt}
{|\vec E|^2}={|E_{ref} + E_b|^2}\cong
{I}{(z,r,\theta,t )} \sim
&& \nonumber \\
{[ 1 + \gamma [2 L_{_{PCM}}- 2 L_{tor}] \cdot {\:}cos[ {\:} (\omega_f-\omega_b ) t + 
{2\ell}{\:}\theta ]]} \cdot {\:}
&& \nonumber \\
{({r}/{D_0})^{2|\ell|}}{\:}
exp{\:}[{\:}-
{\frac {2 r^2}{{D_0}^2(1+z^2/({k_{(f,b)}}^2 {D_0}^4))}}]
{\:}{\:},{\:}{\:}
\end{eqnarray}  
where $z$ is negative provided finite optical thickness of BS 
is neglected (see fig.2). 
For the $\ell$ charged vortices 
\cite { Allen:1992} 
the $2 \ell$ spot output pattern will rotate 
around the common center with the angular velocity 
$\delta \omega/2\ell$  
provided internal PCM mechanism is static and the 
moving internal acoustic or thermal 
waves are absent \cite {Okulov:2012josa,Okulov:2008,Okulov:2010josa}.
The similar interference pattern occurs in the Mach-Zehnder vortex 
interferometer used for excitation of the coherent vortex superpositions 
in quantum gases \cite {Dowling:2005}, 
slow-light media and exciton-polariton condensates \cite {Dowling:2016}.

Formula (\ref {pines_inter_patt}) explains $2 \ell$ spot 
output given by the 
overlapping of the two aligned $LG_{0\ell}$ optical vortices with parallel 
linear and $antiparallel$ angular momenta. 
As in conventional Michelson interferometer 
the visibility of interference pattern at the output port 
of BS is maximal when both arms have equal 
optical length $L_{_{PCM}}= L_{tor}$. The 
experiments of Michelson \cite {Michelson:1904} white light source 
have shown that 
constructive interference at the output port 
occurs even in the case when $L_{_{PCM}}, L_{tor} >> L_{coh}$. 
In our case both helical and toroidal interference patterns 
will also vanish in 
the vicinity of BS (i.e. at small $positive$ $z,z^{'}$). 
Analogously to Michelson white light interferometer the  
output pattern (\ref {pines_inter_patt}) 
will $not$ $be$ $affected$ by finite coherence 
$L_{coh}$ of the source for $L_{_{PCM}} \sim L_{tor}$ 
\cite {Rytov:1987}.

Noteworthy the interference pattern in the nonphase-conjugating arm 
of interferometer located between BS and the reference 
mirror M is composed of the equispaced toroids separated by interval 
$\lambda/2$  \cite {Okulov:2008}:
\begin{eqnarray}
\label{toroidal_inter_patt}
{|\vec E|^2}={|E_f + E_{ref}|^2}\cong
{I}{(z^{'},r,\theta,t )} \sim
&& \nonumber \\
{[ 1 + \gamma [2(L_{tor}-z^{'})] \cdot  {\:}cos[ {\:} \delta \omega_{tor} t - 
(k_f+k_b) z^{'} ]]} \cdot {\:}
&& \nonumber \\
{({r}/{D_0})^{2|\ell|}}{\:}
exp{\:}[{\:}-
{\frac {2 r^2}{{D_0}^2(1+{z^{'}}^2/({k_{(f,b)}}^2 {D_0}^4))}}]
{\:}{\:},{\:}{\:}
\end{eqnarray} 
where $L_{tor}$ is the length of toroidal arm,
$z^{'}$ coordinate originating at beamsplitter BS and 
terminating at mirror M. The 
nonzero $\delta \omega_{tor}$ frequency shift 
is possible in this arm due to OAM tilt in reflections. 
For the $\ell$ charged vortices 
the pattern in toroidal arm is rotationally invariant. 

\section{Geometric phase and the angular Doppler shift accumulation}
Apparently the  angular momentum of photons is not changed 
when viewed 
from reference frames rotating with different angular 
velocities \cite {Landau:1982,Akhiezer:1965}. But when photons 
pass through rotating medium the optical 
angular momentum changes and rotating medium experiences 
recoil \cite {Okulov:2012josa,Fiona:2014}. 
This happens due to isotropy of space \cite {Okulov:2008}. 
Equation (\ref {freq_shift_photon_PC_rot}) 
is valid due to invariance of the Lagrangian of the 
system $photon$ $plus$ $rotating$ $object$ 
with respect to infinitesimal rotations $\delta \theta$. For this reason 
the $ideal$ phase-conjugating mirror will $inevitably$ modify the carrier 
frequency of the reflected PC photons $\omega_b$ \cite {Okulov:2008} for $any$ 
angular speed $\Omega_{\oplus}$ alike Earth 
rotation rate $\Omega_{\oplus} \sim 10^{-5} rad/sec$ and 
even smaller ones. 

The elementary approach based upon conservation 
of energy and angular momentum demonstrated by Dholakia 
\cite {Dholakia:2002} and confirmed in other works 
\cite {Okulov:2012josa} gives also the exact formula for 
the rotational Doppler shift induced by rotation 
of the PCM around propagation axis $\vec Z$: 
\begin{equation}
\label {freq_shift_photon_PC_rot} 
\delta \omega = \omega_b - \omega_f = \pm 
2 \ell {\:} \Omega_{\oplus} + \frac{2 \ell^2 \cdot \hbar}{I_{zz}}, 
\end{equation}  
where $I_{zz}$ is the moment of inertia of PCM 
with respect to $\vec Z$-axis. The second term in the right-hand 
size of (\ref {freq_shift_photon_PC_rot}) is  
negligible for typical masses ($m \sim g$) and 
sizes ($r \sim cm$) of a prisms and mirrors  
$\hbar /I_{zz} \sim \hbar 
/ (m \cdot r^{-2}) \cong 10^{-27} rad/sec$. The frequency 
shift $\delta \omega $ is due to the inversion 
of the angular momentum in reflection 
from rotating PCM ($2\ell \hbar$) and the double passage through 
rotating Dove prism ($4 \ell \hbar)$. 
Using this physically transparent arguments 
\cite {Okulov:2012josa} it is easy   
to obtain expression 
for the $net$ frequency shift for the photon, 
which passed twice, in forward and backward directions, 
through N image inverting elements, say Dove 
prisms \cite {Soskin:2008} 
after reflection from the phase-conjugating mirror  
$\delta \omega_{_{\Sigma}}=4 {\:} \Omega_{\oplus} {\:}  \ell (N+1/2)$.

\subsection{Invariance of the rotational frequency shift}

In a $rest$ (nonrotating)
$frame$ the conservation of energy and angular 
momentum is as follows :

\begin{eqnarray}
\label {vectorial conservation OAM}
\hbar  \omega_f + \frac {|{\vec {\mathcal L}}_{pc}(t)|^2} {2 I_{zz}}
= \hbar \omega_b + \frac {|{\vec {\mathcal L^{'}}}_{pc}(t)|^2} 
{2 I_{zz}} , 
&& \nonumber \\
 {\vec {\mathcal L}}_{pc}(t) +  \vec {{\mathcal L}} =  
 {\vec {\mathcal L^{'}}_{pc}}(t) + {\vec {\mathcal L^{'}}} ,
\end{eqnarray} 
where ${\vec {\mathcal L}}_{pc}(t)$ and ${\vec {\mathcal L^{'}}}_{pc}(t)$ are 
the angular momenta of PCM and 
photons correspondingly 
before $\vec {{\mathcal L}}$ 
and after $\vec {{\mathcal L^{'}}}$ 
photons reflection \cite {Okulov:2012josa}. 
The important simplification is due to the large mass of 
PCM. Let us consider the tensor of inertia $I_{ij}$ 
as indistinguishable from those of spherical 
body ($I_{zz}=I_{yy}=I_{xx}$). Then angular 
momentum $\vec {\mathcal L}$ with respect to 
$arbitrarily$ oriented axis $\vec Z$ is just 
$\vec {\mathcal L}= \vec \Omega_z I_{zz}$. 
Hence in our case (fig.2)
${\vec {\mathcal L}}_{pc}= \vec \Omega_{\oplus} I_{zz}$. 

The slow rotation 
of the frame leads to the perpetual adiabatic tilt of axis 
${\vec Z}(t)||{\vec {\mathcal L}}_{pc}||
{\vec {{\mathcal L^{'}}}_{pc}}$ 
(fig.1). 
Noteworthy the accurate handling 
with angular momentum of 
photons as a classical (!) vector $\vec {{\mathcal L}}{\:}$ 
is compatible with exact quantum picture. Let us  
consider the time-dependent axis 
of photons propagation $\vec Z(t)$ as 
a $measurement$ $axis$ \cite {Landau:1982,Akhiezer:1965}.

In quantum mechanics the projection of 
$\vec {{\mathcal L}}$ on a measurement axis $\vec Z(t)$ may have discrete 
values only ${\mathcal L}_z = -\ell \hbar, ... + \ell \hbar$. 
The multiplication of both sides of second equation 
in (\ref {vectorial conservation OAM}) 
by ${\vec {{\mathcal L}}_{pc}}$ 
yields: 

\begin{eqnarray}
\label {conservation OAM scalar product}
{\vec {\mathcal L}} \cdot  {\vec {\mathcal L}}_{pc}(t)  + 
|{\vec {\mathcal L}}_{pc}(t)|^2 =
{\vec {\mathcal L^{'}}} \cdot  {\vec {{\mathcal L}}_{pc}}(t)  + 
{\vec {{\mathcal L^{'}}}_{pc}}(t) \cdot {\vec {\mathcal L}}_{pc}(t), 
&& \nonumber \\ 
\ell \hbar  |{\vec {{\mathcal L}}_{pc}}| \cos \phi (t) +  
|{\vec {\mathcal L}}_{pc}|^2 =  
- \ell \hbar  |{\vec {\mathcal L}}_{pc}| \cos \phi (t)+  
|{\vec {\mathcal L}}_{pc}| |{\vec {\mathcal L^{'}}}_{pc}|,
\end{eqnarray}

where quantization of photon's angular momentum projection 
is included explicitly though angular momentum of $measurement$  
$device$ (PCM ) remains a  
classical vector ${\vec {\mathcal L}}_{pc}$. 
The second equation is due to the reversal 
of OAM in reflection from the PCM \cite {Okulov:2008}.

After some algebra with first equation in 
(\ref {vectorial conservation OAM}) and second equation in 
(\ref {conservation OAM scalar product} ):
 
\begin{eqnarray}
\label {algebra}
\ell \hbar  |{\vec {\mathcal L}}_{pc}| \cos \phi (t)+  
|{\vec {\mathcal L}}_{pc}|^2 =  
- \ell \hbar  |{\vec {\mathcal L}}_{pc}| \cos \phi (t)+ 
|{\vec {\mathcal L}}_{pc}| |{\vec {\mathcal L^{'}}}_{pc}|,
&& \nonumber \\
\hbar  \omega_f + \frac {|{{\vec {\mathcal L}}_{pc}}|^2} {2 I_{zz}}
= \hbar \omega_b + \frac {|{\vec {\mathcal L^{'}}}_{pc}|^2}
 {2 I_{zz}} , 
{\:}{\:}{\:}{\:}{\:}{\:}{\:}{\:}
\end{eqnarray}

one may obtain the $tiny$ shift of photons carrier 
frequency $\delta \omega$ for $noncollinear$ vectors 
$\vec Z (t) \cdot \vec \Omega_{\oplus} = 
|\vec \Omega_{\oplus}| \cos \phi (t) $ 
as a result of the $stepwise$ phase-conjugating 
OAM reversal from ${\mathcal L_z}=\pm \ell \hbar$ to 
${\mathcal L^{'}_z}=\mp \ell \hbar$:

\begin{eqnarray}
\label {vectorial freq_shift_photon_PC_rot} 
\hbar \delta \omega (t) =
 \frac {{|{\vec {\mathcal L}}_{pc}|}-{|{\vec {\mathcal L^{'}}}_{pc}|}}
{2 I_{zz}} 2 \ell \hbar \cos \phi (t) =  
&& \nonumber \\
- 2 \ell \hbar |\vec \Omega_{\oplus}| \cos \phi (t) 
+ \frac {2 \ell^2 \hbar \cos^2 \phi (t)}{I_{zz}}. 
\end{eqnarray}  
 
This $\delta \omega (t)$ shift coincides 
exactly with (\ref {Coriolis shift}): 

\begin{equation} 
\label {PCM Coriolis shift} 
\delta \omega (t) = 
-  \Delta \vec {\mathcal L} (t) \cdot \vec \Omega_{\oplus}/ \hbar  , 
{\:}{\:}\delta \omega = 
-2 \ell \vec Z (t) \cdot \vec \Omega_{\oplus} ,
{\:}{\:}
|\Delta \vec {\mathcal L} | = 2 \ell \hbar . 
\end{equation}

Hence exchange of the angular momenta between photons and 
tilted rotating vortex interferometer results in 
angular Doppler shift affected by Coriolis multipliers 
($\cos \phi (t)$ and $\cos \xi (t)$) in 
scalar product $\vec Z (t) \cdot \vec \Omega_{\oplus}$. 
As in the case of Foucault pendulum the $2 \ell$ 
spot interference pattern follows to the rotation 
of reference frame $\vec \Omega_{\oplus}$. The angle 
of rotation $\alpha(t)$ in a given moment $t$
equals to the geometrical phase acquired 
by rotating top moving on the surface of a sphere:
 \begin{equation}
\label {geom phase} 
\alpha(t) = - \int_{t0}^{{\:}t} \frac {\Delta \vec {\mathcal L} (t)
\cdot \vec \Omega_{\oplus}}
{ \hbar} {\:} dt 
{\:}{\:}= - 2 \ell  \int_{t0}^{{\:}t} \vec Z (t)\cdot \vec \Omega_{\oplus}  {\:} dt
{\:}{\:}. 
\end{equation}

In $rotating$ 
$frame$ the energy and angular 
momentum conservation for vectorial case is as follows :

\begin{eqnarray}
\label {rot_frame_vectorial conservation OAM}
\hbar  \omega_f + 0 - \ell \hbar \Omega_{\oplus}
= \hbar \omega_b + \frac {|{\vec {\mathcal L^{'}}}_{pc}(t)|^2} {2 I_{zz}} 
+ \ell \hbar \Omega_{\oplus}, 
&& \nonumber \\
{\vec {\mathcal L}} +  0 =  
 {\vec {\mathcal L^{'}}_{pc}}(t) + {\vec {\mathcal L^{'}}} ,
\end{eqnarray} 
 
where $\mp \ell \hbar \Omega_{\oplus}$ is the energy transformation 
due to frame rotation \cite {Landau:1976}. Noteworthy the 
alternation of sign of this term due to reflection 
from PCM \cite {Okulov:2008}. Again after a careful 
algebra the frequency shift viewed in $rotating$ $frame$ will 
be identical to those in $rest$ $frame$:

\begin{equation}
\label {rds in rot frame} 
\delta \omega (t) = 
- 2 \ell {\:}  \vec Z (t)\cdot \vec \Omega_{\oplus}  + 
\frac {2 \ell^2 \hbar (\vec Z (t)\cdot \vec \Omega_{\oplus})^2 }{I_{zz}} 
{\:}{\:}. 
\end{equation}
 
\subsection {Accumulation of rotational Doppler shift }
To accumulate the rotational Doppler shift 
the adjacent image inverting elements (say Dove prisms) 
should rotate in opposite directions. 
This feature is due to vectorial nature of 
angular momentum exchange between photons 
and rotating Dove prisms and PCM. The following "$hand$ $rule$" 
is valid due to 
$\delta \omega = - \Delta \vec {\mathcal L} \cdot \vec \Omega_{\oplus} /{\hbar} $ 
: when angular momenta of photons and 
image inverting element are anti-parallel the energy is transferred to 
photon otherwise rotation of setup is accelerated at the expense 
of photon \cite{Okulov:2012josa}.  
For this reason the $accumulation$ 
effect is algebraical $addition$ not $multiplication$. The 
$accumulated$ frequency shift $\delta \omega_{_{\Sigma}} (z)$ is the stepwise 
function of $z$ (fig.2). The 
smallest speed of the helix rotation  
$|\vec \Omega_{\oplus}|=\delta \omega / 2\ell$ is in 
between PCM and the 
first Dove prism the largest one $\delta \omega_{_{\Sigma}}$ 
is in between the 
last Dove prism and beamsplitter BS. 

In experimental realization the angular speeds of rotation $\vec \Omega_i$ of the 
all $N$ image inverting elements cannot be equal to each other and some 
random spread of angular velocities 
is inevitable: $\vec \Omega_i=\vec \Omega(-1)^{i}+\delta {\vec \Omega}_i$. 
Thus generalization of $\delta \omega_{_{\Sigma}}$ is required 
to include the random 
spread of the rotation frequencies $\delta\Omega_i$: 
\begin{equation} 
\label {random freq_shift_photon_PC_rot} 
\delta \omega_{_{\Sigma}}= -\vec Z \cdot (2 \ell{\:} {\vec \Omega}_{\oplus} + 4 \ell {\:}
\sum_{i=1}^N {\vec \Omega_i}(-1)^{i} {\:} + 
4 \ell {\:} \sum_{i=1}^N \delta {\vec \Omega_i}) , 
\end{equation}
where $\Omega_{\oplus}$ is the angular velocity of PC mirror.
 
Once PC mirror with 
sufficient quality is constructed the $2 \ell$ spot interference pattern 
(fig.2) will 
make one revolution per $86400 \cdot /(2N+1)$ seconds with 
$z$ axis oriented parallel to Earth rotation axis, e.g. in a setup 
located at equator and placed on a 
horizontal optical table.

\subsection{Geometric phase acquired by Foucault pendulum 
and vortex interferometer}

Coriolis force 
$\vec F_c (t)=-2 M \vec \Omega_{\oplus} \times \vec V (t)$,
where $\vec V (t)$ is velocity of Foucault pendulum bob 
in rotating frame, causes 
slow rotation of swinging plane with angular 
frequency $\Omega_{bob} = -|\vec \Omega_{\oplus} | \cos (\phi)$, 
where $\chi=\pi/2 - \phi$ 
is geographical latitude. This follows from Newtonian dynamics 
of the harmonic pendulum 
with mass $M$ 
suspended in slowly rotating frame \cite {Landau:1976}. The 
equation of motion for the bob in rotating frame is:
\begin{equation}
\label {Newtonian pendulum Coriolis vectorial}
M {\vec a} = \vec F_{g} + \vec F_{c} = 
- M \vec g - M 2[\vec \Omega_{\oplus} \times \vec V],
\end{equation} 
where $M \vec g$ is the local gravity force. For the small 
amplitude oscillations in the $(x,y)$ plane, where the $y-axis$ is 
North directed and the $x-axis$ is East directed at 
$\chi$ geographical latitude, the coupled equations 
for harmonic oscillations are: 

\begin{eqnarray}
\label {Newtonian pendulum Coriolis scalar}
 \ddot x = -\omega^2 x + 2 \Omega_{\oplus} \cdot \dot y \cdot  \sin \chi ,
&& \nonumber \\
 \ddot y = - \omega^2 y - 2 \Omega_{\oplus} \cdot \dot x \cdot  \sin \chi ,
\end{eqnarray} 
where $\omega = \sqrt {|\vec g| {L_{F}}} = 2 \pi / T_{F} $ is 
angular frequency of bob oscillations, 
$L_{F}$ is length of suspension wire.
For complex vector $z=x+iy$ this 
system becomes:

\begin{equation}
\label {Newtonian pendulum Coriolis complex plane}
\frac {d^2 z}{d t^2} + 
2 i \Omega_{\oplus} \cdot \frac {d z}{d t} \cdot  \sin \chi   + \omega^2 z=0 ,
\end{equation} 
with obvious solution to the first order 
in $\Omega_{\oplus}/ \omega$: 
 
\begin{equation}
\label {solution Newtonian pendulum Coriolis complex plane}
z = \exp [- i \Omega_{\oplus}  \sin \chi {\:} t] [c_1 \exp (i \omega t)
+c_2 \exp (-i \omega t)] ,
\end{equation}
where the arbitrary constants $c_1,c_2$ come from initial conditions.
Noteworthy the absence of the rest mass $M$ here. 
Thus swing plane of the Foucault pendulum 
rotates with angular velocity 
$\Omega_{F}=-\Omega_{\oplus} \sin \chi $ around local gravity 
acceleration vector $\vec g$.  
Apparently the modulus of 
geometrical phase $\alpha = -2 \pi \sin \chi $ acquired during 
one rotation 
reaches the maximal value at the Poles (fig.1). 

For the observer in 
a reference frame standing on the Earth the trajectory of 
$bob$ becomes is curvilinear due to Coriolis force 
with the time dependent angular momentum 
$\vec {\mathcal L}_{F} (t)$ directed along $-\vec g$: 
\begin{equation}
\label {Angular momentum Foucault definition}
{\mathcal {\vec L}}_{F}(t) =  [\vec { r} \times \vec {p}]=
M [\vec { z} \times \dot {\vec z}],  
\dot {\vec z} + \vec \Omega_{\oplus} \times \vec r = \vec V . 
\end{equation}
After some algebra the angular momentum projection 
on $- \vec g$ as a function of time ${\mathcal { L}}_{fouc}(t)$ 
might be obtained under zero velocity initial condition
$\dot {\vec z}=0$ for $t=0$ : 
\begin{eqnarray}
\label {Angular momentum Foucault result}
{\mathcal { L}}_{F}(t) =  
M \cdot [- 2 c_1 c_2 \Omega_{\oplus}\sin \chi  \cos (2 \omega t) +
&& \nonumber \\
(c_1 - c_2)(c_1 + c_2) \omega \cos (\Omega_{\oplus} \sin \chi {\:}   t)^2 \cong  
&& \nonumber \\
- M  \Omega_{\oplus} \sin \chi \cdot 2 c_1 c_2 \cdot \cos (2 \omega t) , 
{\:} {\:}\leftarrow {\:} {\:} \Omega_{\oplus} << \omega. 
\end{eqnarray}  

Hence angular momentum  ${\mathcal {L}}_{F} (t)$ 
oscillates with period $\pi / \omega = T_{F}/2$.

In Michelson vortex interferometer (fig.2) 
the photons with zero rest mass 
and the angular momentum ${\mathcal L}_{Z}=\pm \ell \hbar$ 
are also affected by frame rotation when 
their angular momentum direction is changed via 
phase conjugation and passage through Dove prisms 
\cite {Okulov:2012josa}. The optomechanics of this 
$spin-$ $orbital$ interaction is a $\delta$ - kicked one: 
the most of the time 
the photons with angular momentum $ {\mathcal L}_{z} = \pm \ell \hbar$ 
moves in free space. At the moments separated by time of flight 
intervals 
$2 \Delta L_{_{pc}}/c$ the $\delta$-kicks adjust the helical 
phase front $\exp (i \ell \theta)$ to the gradually changing 
orientation of interference pattern inside PCM. 

The reversal of the photons angular momentum in rotating 
PCM and Dove prism might be interpreted as effective 
"Coriolis" force induced by a slowly moving fringes of 
interference pattern in PCM and tilted planes in Dove prism. 
This leads to rotation of the helical interference pattern 
\cite {Okulov:2012josa} with the the similar angular velocity 
$\Omega_{_{pc}} = \delta \omega /2 \ell =
- |\vec \Omega_{\oplus} | \cos \phi =  \Omega_{F} $
as it happens with Foucault pendulum. 

In both cases the initial conditions are essential. 
In PCVI (fig.2) the interference fringes inside PCM and 
vortex fringes 
must be adjusted when holographic plate is used as PCM , 
while Foucault pendulum $bob$ ought to be 
gently released 
from maximal 
deflection point 
with zero initial velocity $\dot {\vec z}=0$. Then $bob$ begins to 
fall towards equilibrium 
position but its trajectory gradually bends because 
of Coriolis force which is the source of the periodically modulated angular 
momentum $ {\mathcal {\vec L}}_{F}(t)$ \cite {Foucault:1852} .

\subsection{The mutual orientation of the setup 
rotation $axis$ $\vec \Omega_{\oplus}$ and vortex propagation 
$axes$ $\vec Z$ and $\vec Z'$ }

Consider the important issue which stems from the angular momentum 
transformation in PCVI. 
Noteworthy the case when axis of rotation $\vec \Omega_{\oplus}$ 
and $toroidal$ axis $\vec Z'$ are mutually orthogonal the net 
angular Doppler shift 
is absent ($\delta \omega=0$). 
In this case backward wave in $toroidal$ arm acquires the 
additional Doppler shift $\delta \omega_{tor}=\ell {\:}  \Omega_{\oplus}$ due to 
angular momentum tilt at $90^{o}$ in BS after backward 
reflection from BS. 
This case is the worst suited  
for usage as reference wave to observe the beats 
(\ref {pines_inter_patt}) with 
angular frequency $\delta \omega$ 
due to superposition with backward wave from $helical$  arm. 
For orthogonal helical and toroidal arms the rotational Doppler shift 
between waves in the output port 
is exactly zero. This follows from $hand$ $rule$ 
used for analysis of OAM transformation in passage 
through BS, PCM
and reflection from M. The best mutual orientation 
of $\vec Z$ and $\vec Z'$ is 
to be almost parallel ($\eta \rightarrow 0$) 
in order to minimize the OAM change via deflection inside 
BS because the later rotates together with setup. 
For this reason the different tuning angles are selected at (fig.2): 
$\phi$ is the angle between rotation axis $\vec \Omega_{\oplus}$ 
(say targeted to Polar star) 
and helical (PCM) axis, while $\eta$ is angle between 
$\vec Z$ and $\vec Z'$ (\ref {toroidal_inter_patt}).

\section {Vortex Michelson interferometer and 
the laser noise}
\subsection {Phase noise and visibility of helical pattern}

The fundamental limit on the laser phase noise is given 
by the Shawlov-Townes 
formula \cite {Townes:1958} which connects the laser linewidth 
$\delta \nu_{_{ST}}$ with 
emitted power $P$, cavity mode bandwidth $\Delta \nu_c$ 
and effective temperature 
of the lasing medium $T_L$: 
\begin{equation}
\label{Townes_limit}
\delta \nu_{_{ST}} \sim 
4 \pi (\hbar \omega_{f,b}+k_B T_L ) (\Delta \nu)^2 /P . 
\end{equation}
The amplitude fluctuations are assumed to form 
background of the narrow 
stimulated emission line $\delta \nu_{_{ST}} $ 
which is due to the phase fluctuations 
diffusion with characteristic coherence time $\tau_c$. 

In order to generalize the previous analysis \cite{Dholakia:2002D} 
for visibility of patterns in PC vortex 
interferometer \cite {Okulov:2008, Woerdemann:2009} (fig.1),
let us take for 
definiteness the electric field envelopes in the form 
of Laguerre-Gaussian beam (LG) \cite {Okulov:2008}: 

\begin{eqnarray}
\label{pump1}
{{E}_{(f,b)}(\vec r,t)}
\sim {\frac {\exp [ -i \omega_{(f,b)}t 
\pm i k_{(f,b)} z \pm i{\ell}\theta + i \Theta_{(f,b)}(t)]} 
{ {(1{+}iz/z_R)}} } 
&& \nonumber \\
{{\ E}_{(f,b)}^0}{({r}/{D_0})^{|\ell|}}
\exp  [  - 
{\frac {r^2}{{D_0}^2(1{+}iz/z_R)}} ], 
z_R=k_{(f,b)} {D_0}^2{\:} {\:} 
\end{eqnarray}
or a Bessel beam optical vortex \cite {Sepulveda:2009}:
\begin{eqnarray}
\label{pump2}
{\bf{E}_{(f,b)}(\vec r,t)}
\sim{{\ E}_{(f,b)}^0} \cdot J_m(\kappa r)
&& \nonumber \\
{  \exp [ -i \omega_{(f,b)}t 
\pm i k_{(f,b)}z \pm i{\ell}\theta + i \Theta_{(f,b)}(t)]},
\end{eqnarray}
where the cylindrical coordinates $\vec r=(z,r,\theta)$ 
are used,  
$\bf{E}_f$ stands for the forward 
wave (fig.1), propagating in positive Z-direction, 
$\bf{E}_b$ stands for the wave, propagating in 
the opposite direction , $J_m(\kappa r)$ 
is the m-th order Bessel function, $\Theta_{(f,b)}(t)$ are the 
random variations of the phases of partial waves, 
induced by both the laser linewidth and fluctuations inside  
interferometer \cite {Siegman:1986}. 

The interference pattern produced by two vortices 
with phase noise $\Theta_{f}(t)$ and $\Theta_{b}(t)$ 
is as follows:

\begin{eqnarray}
\label{noisy_helical_inter_patt}
{|\vec E|^2}={|E_{f} + E_b|^2} \cong
I(z,r,\theta,t ) \sim 
&& \nonumber \\
{[1 + [cos (\omega_f-\omega_b ) t-(k_f+k_b)z + 
{2\ell}{\:}\theta +\Theta_{f}(t)-\Theta_{b}(t)]]} \cdot {\:}
&& \nonumber \\
{({r}/{D_0})^{2|\ell|}}{\:}
exp{\:}[{\:}-
{\frac {2 r^2}{{D_0}^2(1+z^2/({k_{(f,b)}}^2 {D_0}^4))}}]
{\:}{\:}{\:}{\:}{\:}{\:},{\:}{\:}
\end{eqnarray} 

The pattern breaths due to slow random drift of phases 
$\Theta_{f}(t)$ and $\Theta_{b}(t)$. Taking into 
account that counterpropagating vortices are 
produced via reflection from PCM, i.e.
$\Theta_{f}(t)=\Theta(t+T)=\Theta_{b}(t-T)$ 
where $T=2(L_{PCM}-z)/c$ is time delay 
at a given point $z$ on interferometer axis, 
we have for  statistically averaged interference pattern 
at output port (fig.2): 
\begin{eqnarray}
\label{averaged_noisy_helical_inter_patt}
\overline{|E_{f} + E_b|^2} \sim 
\lbrace 1 + {\:}[1- \overline {\frac 
{(\Theta(t)-\Theta(t-T))^2}{2!}}]{\:} 
{\:}{\:} {\:}{\:} {\:}
&& \nonumber \\
 \cdot cos[ {\:} 
(\omega_f-\omega_b ) t -(k_f+k_b)z + 
{2\ell}{\:}\theta ] \rbrace \cdot {\:}
&& \nonumber \\
{({r}/{D_0})^{2|\ell|}}{\:}
exp{\:}[{\:}-
{\frac {2 r^2}{{D_0}^2(1+z^2/({k_{(f,b)}}^2 {D_0}^4))}}]
{\:}{\:}{\:}{\:}{\:}{\:},{\:}{\:}
\end{eqnarray}  
with apparent averaging of the first term 
$\overline {(\Theta(t)-\Theta(t-T))}=0$ 
of this Taylor expansion justified for 
narrow linewidth $\delta \nu_{_{ST}}$.
The second order correlation function $\gamma$ or visibility 
of pattern is as follows:

\begin{equation}
\label{visibility due to phase noise}
\gamma [2(L_{PCM}-z)] = 1- \overline {\frac 
{(\Theta(t)-\Theta(t-T))^2}{2!}} ,
\end{equation}
 with apparent 
limit $\gamma \rightarrow 1$ as $T\rightarrow 0$.

\subsection {Quantum state transformation 
in the ideal vortex phase-conjugator}
\begin{figure} 
\center{ \includegraphics[width=7.0 cm]{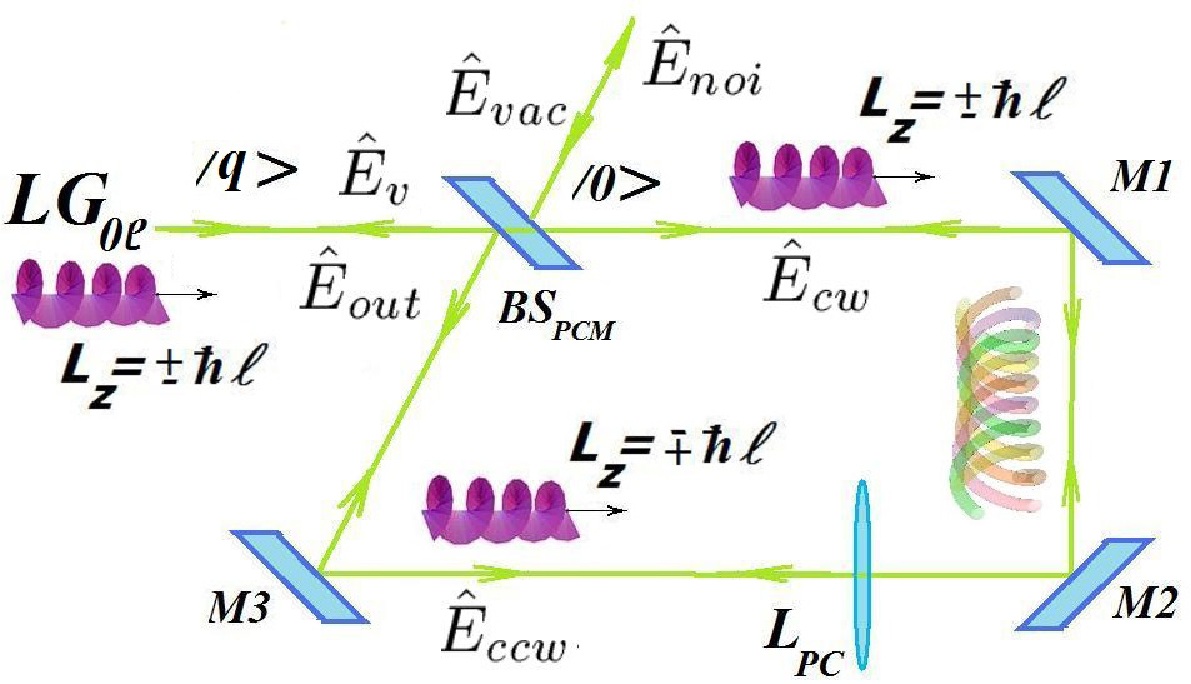}}
\vspace {2mm}
\caption{(Color online) Ideal PCM for isolated vortex photons.
$BS_{_{PCM}}$ is entrance beamsplitter of PCM, $M_{1-3}$ are 
conventional mirrors, $L_{PC} $ is wavefront 
matching lens, $|0>$ is vector of state of vacuum, $|q>$ 
is a vector of state 
of vortex photon emitted by laser.}
\label{fig.3}
\end{figure}
Generally speaking statistics of the phase-conjugated photons 
\cite {Okulov:1983} changes due to the mixing with amplified 
vacuum modes\cite {Caves:1981} incident from 
open rear ports of PCM \cite {Gaeta:1988}. However 
there exists an example of the PCM that does not 
change statistics of PC-reflected photons. Indeed the Sagnac loop 
PCM proposed in \cite {Okulov:2010josa} (fig.3)
may operate without coupling of incident signal with 
external modes in the case of negligible absorption 
and amplification inside PCM. This happen in a quite realistic 
case of the perfect 50/50 beamsplitter when each incident vortex 
photon in an $arbitrary$ quantum state $|q>$ moves along two 
paths (clockwise and counterclockwise) with identical optical 
lengths $L_{+}$ and  $L_{-}$. The same happens with the vacuum 
modes $|0>$ incident through the open port of $BS_{_{PCM}}$ and 
recombining after propagation along the 
equal paths $L_{+}$ and  $L_{-}$. In Heisenberg picture 
the interference pattern inside PCM looks 
as follows:
 
\begin{eqnarray}
\label{Heisenberg operators transformation}
{\hat E}_{ccw} = \sqrt{R_{in}} {\:}{\hat E}_{v} + 
\sqrt{T_{in}} {\:}{\hat E}_{vac},
&& \nonumber \\
\hat E_{cw} = \sqrt{T_{in}} {\:}\hat E_{v} - \sqrt{R_{in}} {\:}\hat E_{vac},
\end{eqnarray}
 
where $\hat E_{v}$ is the incident vortex field, 
$\hat E_{vac}$ is the vacuum zero-point field, $\hat E_{cw}$ is 
clockwise field inside the PCM, $\hat E_{ccw}$ is 
counterclockwise field inside the PCM, $R_{in}$ and $T_{in}$ are 
intensity reflectance 
and transmittance of the entrance beamsplitter 
$BS_{_{PCM}}$ of PCM. Then output 
annihilation (electric field) operators $\hat E_{out}$ and 
 $\hat E_{noi}$ are:

\begin{eqnarray}
\label{Heisenberg operators transformation_2}
\hat E_{out} = \sqrt{T_{in}} {\:}\hat E_{ccw} + 
\sqrt{R_{in}} {\:}{\hat E}_{cw},
&& \nonumber \\
\hat E_{noi} = \sqrt{T_{in}} {\:}\hat E_{cw} - \sqrt{R_{in}} {\:}\hat E_{ccw},
\end{eqnarray}

or in the terms of the input fields operators:

\begin{eqnarray}
\label{Heisenberg operators transformation_3}
\hat E_{out} = 2 \sqrt{T_{in}R_{in}} {\:}\hat E_{v} + 
(T_{in}-R_{in}) {\:}{\hat E}_{vac},
&& \nonumber \\
\hat E_{noi} = -2 \sqrt{T_{in}R_{in}} 
{\:}\hat E_{vac} + (T_{in}-R_{in}) {\:}\hat E_{v}.
\end{eqnarray}

Thus for peculiar case $T_{in}=R_{in}=1/2$ the noise 
contribution in the output field 
$\hat E_{out}$ is completely suppressed and the incident vortex photons 
return via constructive interference at $BS_{_{PCM}}$ 
in exactly opposite direction with inverted 
momentum $ - \vec k$ 
and opposite angular momentum $- \vec L$. 
As a result the all zero-point 
noise incident through the open rear port  $BS_{_{PCM}}$ 
is removed via constructive interference 
at $BS_{_{PCM}}$. 

The composite quantum state of the vortex photons and 
vacuum fluctuations 
 $ |\psi>=|q> |0> $ remains unperturbed in Heisenberg picture. 
Moreover for  $T_{in}=R_{in}=1/2$ the statistics is not changed. 
The wavefront matching lens 
$L_{_{PC}}$ located at the midway of the loop (fig.3) compensates 
the natural beam divergence of vortex and external 
vacuum fields. This remove the  
mixing with another propagation modes different from 
incident vortex.

The vortex linewidth may be evaluated  
as coherent state $CS$ whose linewidth 
$\delta \nu_{_{CS}} \lesssim  {\:}{\:}\delta \nu_{_{ST}} $ 
approximately corresponds to uncertainty relation for number 
of photons $n$ and phase $\phi$ \cite{Dowling:1998}: 
 
\begin{equation} 
\label{coherent_state_limit}
{\:}{\:}{\:}
\delta n \cdot \delta \phi \sim 1;  \delta \phi = \frac {1}{\delta n} = 
\frac {1}{\sqrt n} = \sqrt {\frac {\hbar \omega_{(f,b)}}{\epsilon_o V I} }, 
\end{equation}
 where $\epsilon_0$ is free space permittivity, $V$ is mode volume, 
$I \cong |E_{(f,b)}|^2$ is light intensity . 
One may suggest that a higher density of the interference 
fringes ($2 \ell$ per $\lambda/2$) in helical 
interference pattern may enhance the  
phase sensitivity of Michelson interferometer 
by a factor $(2\ell) ^ \alpha$ \cite{Lugiato:1997}, 
where $1 \le \alpha \le 2$: 

\begin{equation} 
\label{structured_coherent_state_limit}
{\:}{\:}{\:}
(2\ell ) ^ \alpha \delta n \cdot \delta \phi \sim 1; 
 \delta \phi = 
\frac {1}{(2 \ell) ^ {\alpha} \sqrt n} =
\sqrt {\frac {\hbar \omega_{(f,b)}}
{\epsilon_o V I(2 \ell) ^ {2\alpha}} }.
\end{equation}
Furthermore in a structured squeezed state limit 
\cite{Dowling:2008,Lugiato:1992} the ultimate 
resolution may reach even the smaller values: 
 
\begin{equation} 
\label{structured_squeezed_state_limit}
{\:}{\:}{\:}
(2\ell ) ^ \alpha \delta n \cdot \delta \phi \sim 1; 
 \delta \phi = \frac {1}{\delta n} =  
\frac {1}{ n} = (2 \ell) ^ {-\alpha} 
\frac {\hbar \omega_{(f,b)}}{\epsilon_0 V I} . 
\end{equation}

Such a significant reduction of the phase uncertainty requires the 
accurate quantum measurements procedures \cite {Haus:1986}.
In this way the sensitivity of Michelson interferometer to coarse 
displacements might be improved by $3 \div 4$ orders of magnitude
using currently achievable vortex laser beams with topological 
charges $\ell \sim 10^3 \div 10^4$ \cite {Zeilinger:2015}. 
The underlying idea is that 
the structured interference pattern with $2 \ell$ spots 
behind output beamsplitter (fig.2) 
will rotate faster when $\ell$ is increased. Though the phase fluctuations 
of the laser beam 
could mask the external perturbations of space-time metric 
(optical path difference in fig.2 ) the coarse azimuthal 
interference pattern structured by the $2 \ell$ factor 
would facilitate the detection of tiny 
changes of optical paths. 
This conjecture requires the detailed quantitative 
study with the quantum measurements theory\cite {Scully:1997}. 

\vspace {2mm}

\section{Conclusions} 

In summary the Michelson phase-conjugating 
vortex interferometer (fig.2) 
had been analyzed for the purpose of detection of the 
ultraslow rotations and tiny metric disturbances 
using concept of an $ideal$ PCM  
\cite {Okulov:2010josa} and a fact that 
optical vortex propagation in free space is not affected by a choice 
of reference frame \cite {Okulov:2012josa,Fiona:2014}. 
The novel feature compared 
to \cite {Okulov:2008} which is close to Beth's 
$spin$ of photon $phase-conjugating$ torsion pendulum experiment
\cite {Beth:1936} is in the additional reference arm where the nonrotating 
vortex beam is stored \cite {Woerdemann:2009} (fig.2). 
This gives the robustness of the interferometer  
and possibility to use the broadband light source 
with $L_{coh}>|L_{_{PCM}}-L_{tor}|$ in contrast to previous studies 
\cite {Okulov:2012josa,Okulov:2008} where $L_{coh}>L_{_{PCM}}$ is the  
necessary condition for the visible interference pattern. 
The motion of the circular interference 
fringes due to changes of the optical paths difference
resembles the operation of the Foucault 
pendulum \cite {Foucault:1852} where bob marks the points on the circle 
corresponding to a given rotation angle of reference frame. 

The actual range of detectable frequencies of the slow rotations might be
augmented by 
a set of image-inverting counter-rotating elements $N$ in PCM arm of PCVI.
The accumulated angular Doppler shift grows 
linearly with $N$. The explicit 
expression have been obtained for $\delta \omega_{net}$ including  
the random spread of rotation velocities $\Omega_i$. 

In simplest configuration, i.e. without accumulating  
Dove prisms with oppositely directed angular velocities $\vec \Omega_i$,
the proposed vortex interferometer will contain no rotating parts or 
lasers with unique features. The propagation of vortex beams follows here to 
the conventional transformation optics \cite {Korotkova:2022}.

As a well known Beth setup for the optical torque 
measurement \cite {Beth:1936} and Mach-Zehnder 
vortex interferometer 
for the rotational Doppler effect 
demonstration \cite {Dholakia:2002} the our proposal (fig.2) 
is based entirely on the Maxwell equations 
and Einstein relativity without usage of any additional 
assumptions alike "ether theory".  

The ultimate performance of the Michelson vortex interferometer 
is limited by the quantum noise of laser as in conventional 
Michelson interferometer \cite {Caves:1981}. Hopefully there 
exists opportunity to improve the sensitivity by a factor $(2 \ell)^{\alpha}$ 
with the aid of the azimuthally structured interference pattern 
of topologically charged laser beams 
\cite {Okulov:2008,Woerdemann:2009}. 
The further increase of the sensitivity looks promising with the 
superfluid interferometry in lattices \cite{Cuevas:2007},
helical traps \cite{Okulov:2012,Okulov:2013,Lembessis:2016,Lembessis:2017} 
and exciton-polariton condensates \cite {Dowling:2016}.

The mechanism of detection is the $dragging$ of $2 \ell$ spot 
interference pattern formed by interference fringes within PCM. 
In this minimal configuration vortex  
interferometer is the optomechanical 
proof of the 
isotropy of space. 
From the point of view of observer collocated 
with interferometer in the slowly rotating frame 
the $2 \ell$ 
spot pattern circumvents around the LG beam axis. 
To this rotating observer the vortex beam 
reflected from the 
PCM acquires the angular Doppler shift $\delta \omega$. 
On the other hand from the point of view of 
observer placed on "remote unmovable star" (i.e.$rest$ $frame$) the 
anisotropic PC mirror $drags$ the twisted 
interference pattern. 

\section{Disclosures}
Author declare no conflicts of interest.


\begin{thebibliography}{99}
\bibitem{Sagnac:1913}G. Sagnac, "On the proof of 
the reality of the luminiferous aether by the 
experiment with a rotating interferometer",
Comptes Rendus, vol.157, p. 1410-1413 (1913).
\end{thebibliography}
\end{document}